# Convolutional Neural Network and decision support in medical imaging: case study of the recognition of blood cell subtypes

Daouda DIOUF[1], Djibril SECK[2], Mountaga DIOP[2] and Abdoulaye BA[3]

[1]Laboratoire de Traitement de l'Information, Ecole Supérieure Polytechnique, Université cheikh Anta Diop, Dakar, Senegal
[2]Institut national Supérieur de l'Education Populaire et du Sport, Université cheikh Anta Diop, Dakar, Senegal
[3]Faculté de Médecine, de Pharmacie et d'Odonto-Stomatologie, Université cheikh Anta Diop, Dakar, Senegal

**Abstract:** Identifying and characterizing the patient's blood samples is indispen-sable in diagnostics of malignance suspicious. A painstaking and sometime sub-jective task are used in laboratories to manually classify white blood cells. Neural mathematical methods as deep learnings can be very useful in the automated recognition of four (4) subtypes of blood cells for medical application.
The purpose of this study is to use deep learning for image recognition of the four (4) blood cell types and to enable it to tag them. These approaches therefore depend on convolutional neural networks. To do this, we have a dataset of blood cells with labels of the corresponding cell types.
The elements of the database are the input of our convolution which is a simple mathematical tool that is widely used for image processing. These databases have allowed us to create learning models for image recognition, particularly of the blood cell type. Based on the fact that a deep neural network model is able to rec-ognize each element of a scene provided it has been trained for this purpose, this activity focused on carefully selecting the optimization parameters of the model. We evaluated the recognition performance and outputs learned by the networks in order to implement a neural image recognition model capable of distinguishing polynuclear cells (neutrophil and eosinophil) from those of mononuclear cells (lymphocyte and monocyte). The classification accuracy on the learning dataset is 97.39% and the validation accuracy is 97.77%. Images detection failure is very low.

# 1 Introduction

The white blood cell, that is an essential part of the immune system, can be classified into five types as eosinophils, lymphocytes, neutrophils, monocytes and basophils. Microscopic differential white blood cell count is still performed by hematologists, being indispensable in diagnostics with malignance suspicious [1].

By putting blood smear on glass slide and using dyes, it become possible to see cellular structures. This allows to differentiate Red Blood Cells (RBCs) and White Blood Cells (WBCs). It is also possible to detect presence of anisocytosis, blood parasites, and so on [2].

Traditional method consist of the use of microscopic analysis of peripheral blood smear. It is know that this method is costly and time-consuming. A trained medical technician takes about 15 min to evaluate and count 100 cells for each blood slide with susceptible risk error procedure and time consuming [3].

That is why, permanently, researchers develop machine learning algorithms, computer vision, image processing, etc, for automated analysis microscopic blood smear images to improve analysis process and the accurateness.

For classifying blast cells from normal lymphocyte cells, Joshi et al. (2013) used a K-NN classifier. An accuracy rate of 93% is obtained according to the test results. They also proposed the Otsu's automatic thresholding algorithm for segmentation of blood cells. [4]

Tantikitti et al. (2015) used decision tree method to classify, with 92.2% accuracy, 167 cell leukocyte. The also classify, with 72.3% accuracy, 264 blood cells to detect dengue virus infections of patients [5].

Xu et al. (2017) employ a deep convolutional neural networks (CNNs) to classify sickle shape RBCs in an automated manner with high accuracy [6].

Mass processing of the data makes possible to set up a deep learning model for the automatic detection and classification of cells such as eosinophil, lymphocytes, monocytes and neutrophils. Deep learning began with an architecture with masks and shared weights proposed by Yann Le Cun et al in the early 1990s [7]. The shared weight method is equivalent to performing a convolution operation on the inputs. This will be called CNN (Convolutional Neural network). This technique was used for the recognition of handwritten numbers [8]. The convolution operation can be followed by a pooling method, for example by averaging the values of a sub-region or by taking the maximum. With these 2 types of repeated operations, combined and possibly supplemented by others, different layers of non-linear calculation units are created. This defines an architecture that is all the more profound when there are layers. Our work consisted in image recognition of blood cell subtypes using a deep learning architecture that uses in a classical way the minimization of a cost function using the gradient back-propagation algorithm [9]. To do this, we have a data set containing 1600 augmented images of blood cells with labels of the corresponding cell types. There are about 400 images for each of the 4 different cell types grouped together. Cell types are eosinophil, lymphocytes, monocytes and neutrophils. All these images are from the BCCD database, which is a small-scale database for the detection of blood cells. BCCD is under MIT license. From these dataset, we create database including:
- 1600 images for learning (and validation), divided into 4 folders of 400 images each.
- 213 images for the test. The images correspond to 4 classes of blood subtypes: neutrophil, eosinophil, lymphocytes and monocytes.

## 2　Convolutional neural network

Convolutional networks were first introduced by Fukushima. He derived a hierarchical nerve network architecture inspired by Hubel's research work [10]. Lecun generalized them to successfully classify the numbers and to recognize handwritten control numbers. A convolutional neural network consists of several layers [11]. Fig. 2 shows these different layers.

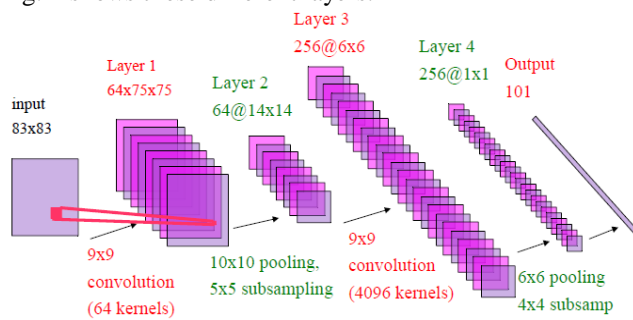

Fig. 1: Example of a convolutional network

*2.1 Convolutional layers*

The convolutional layers constitute the core of the convolutional network. These layers consist of a rectangular grid of neurons that have a small receptive field extended throughout the depth of the input volume. Thus, the convolutional layer is just an image convolution of the previous layer, where the weights specify the convolution filter.

*2.2 Pooling layers*
After each convolutional layer, there may be a pooling layer. The pooling layer under samples their input. There are several ways to do this pooling, such as taking the average or maximum, or a linear combination learned from the neurons in the block. For example, Fig. 1 shows max pooling on a 10×10 window fully connected layers.
Finally, after several layers of convolution and pooling, high-level reasoning in the neural network is done via fully connected layers. In convolutional neural networks, each layer acts as a detection filter for the presence of specific characteristics or patterns present in the original data. The first layers of a convolutional detect characteristics that can be recognized and interpreted relatively easily. Subsequent layers increasingly detect more abstract characteristics. The last layer of the convolutional network is capable of making an ultra-specific classification by combining all the specific characteristics detected by the previous layers in the input data. In the following section, the proposed architecture of the convolutional network is presented.

## 3   Proposed convolutional neural network

*3.1 The structure*
Our classification architecture is standard. It combine convolution and Max pooling. However, to obtain a quick classification allowing real-time classification and localization, we chose a lightweight network. Fig. 2 shows the seven layers of our convolutional network. A color image passes successively through a convolutional operation with a 9x9 nucleus size. The same structure is applied after the third coat. A Max pooling 2x2 with step 2 follows convolutional layers two and four. Layers C2 to C4 have 32 feature maps and layers C5 and C6 have 64 feature maps. Layer C6 and C7 are fully connected. The output of the last layer fully connected feeds a 4-way Softmax producing a distribution over 4 classes. **T**he CNN varies in the way the maximum convolution and pooling layers are realized and in which the layers are formed.

As shown in Fig. 2, the network contains six layers with weights, including the input layer C1, the convolution layer C2, the max pooling layer C3, another convolution layer C4 followed by a second layer of max pooling C5, called full connection, and the output layer C6. Assuming that $\theta$ represents all the parameters that can be trained (weight values), $\theta = \{\theta i\}$ and = 1, 2, 3, 3, 4, 5, 6 where $\theta i$ is the parameter defined between (i- 1)$^{th}$ and i$^{th}$ layer.

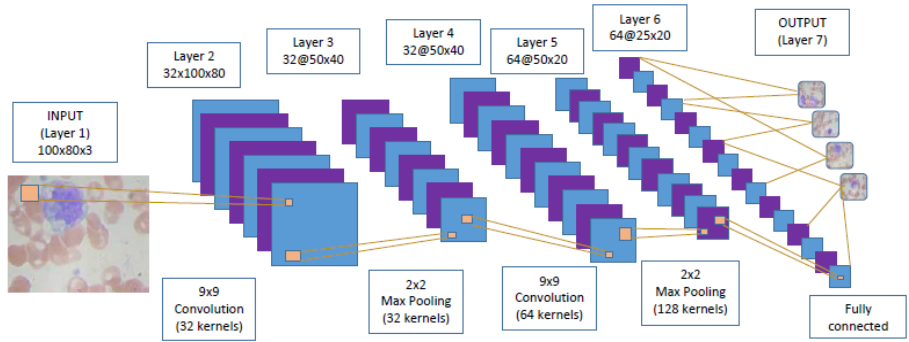

Fig. 2: Architecture of the proposed CNN classifier. The input represents a 2D image, followed by convolution layers and max pooling layers to calculate a set of 32 and then 64 classified feature maps with a fully connected network.

Each sample is considered as a 2D image whose height is equal to 3. Therefore, the size of the input layer is (80, 100) and $n1$ is the number of pixels (i.e. n1=80x100). The first hidden convolution layer C2 filters the $n1$ input data with 32 nuclei of size k1=81 (9x9). The C2 layer contains $32 \times 100 \times 80$ nodes and is of size $n2 = n1$ There are $32\times [(9\ x9)+ 1]$ parameters that can be driven between layer C2 and the input layer. The maximum pooling layer (max pooling) C3 is the second hidden layer and the size of the core is (2, 2) so size k2=4. Layer C3 contains $32 \times 50 \times 40$ nodes and n3 = n2 / $k2$. There are no parameters in this layer. The C4 layer contains $64\times 50 \times 40$ nodes and is sized $n4 = n3$. There are $64\times [(9\ x9)+ 1]$ parameters that can be driven between layer C4 and layer C3. The maximum pooling layer (max pooling) C5 is the fifth hidden layer and the size of the nucleus is (2, 2) therefore of size k4=4. The C5 layer contains $64 \times 25 \times 20$ nodes and n5 = n4 / $k4$. The fully connected layer C6 has n6 nodes and there are $(64 \times[2x2] + 1) \times$ n5 parameters that can be driven between this layer and the layer C5. The C7 output layer has n6 nodes and there are (n6 +1) $\times$ $n7$ parameters that can be formed between this layer and the C6 layer. Therefore, the architecture of our proposed CNNs classifier has a total of $32 \times (9x9 + 1) + 64 \times (9x9 + 1) + 64 \times [2x2] + 1) \times n5 + (n6 + 1) \times n7$ parameters that can be driven.

In our architecture, layers C1 to C5 can be considered as an entity extractor that can be driven for input data, and layer C6 is a classifier that can be driven for the entity extractor. The output of the sub-sampling is the real entity of the original data. In our proposed CNN structure, 32 characteristics can be extracted from each image, and each entity is 50x40 in size.

*3.2 Network training*

The purpose of our classification is to determine whether an image contains the eosinophil, or neutrophil, or monocyte or lymphocyte type. To solve this problem, the classifier's learning is performed from a collection of images in the visible spectrum in 3 labelled channels (RGB). In addition, we want to determine the type of blood in an image.

The training was carried out with a computer composed of an Intel Corei5 microprocessor (CPU frequency 2.7 Ghz, RAM 8Gb).

The learning process consists of two steps: feed-forward propagation and backward propagation. The purpose of forward propagation is to calculate the actual classification result of the input data with the current parameters. Backward propagation is used to update the parameters that can be driven in order to minimize the difference between the actual classification output and the desired classification output.

**3.2.1 Feed-forward propagation**

Our layer (L + 1) of the CNN network (L = 7 in this work) consists of $n_1$ input units in the INPUT layer, $n_5$ output units in the OUTPUT layer and several so-called hidden units in layers C2, C3, C4, C5 and C6. Assuming that $x_i$ is the input of the i[th] layer and the output of the (L-1)[ith] layer, we can calculate as follows:

$$x_{i+1} = f_i(u_i),$$
with
$$u_i = W_i^T x_i + b_i,$$

and $W_i^T$ is a weighting matrix of the i[th] layer acting on the input data and b is an additive bias vector for the i[th] layer and $f_i$ is the activation function of the i[th] layer. In our designed architecture, we have chosen the hyperbolic tangent function *tanh(u)* as the activation function in layers C2, C4 and C6. The maximum max function (u) is used in layers C3 and C5. Since the proposed CNN classifier is a multi-class classifier, the output of layer C6 is transmitted to the 4-way softmax function which produces a distribution on the n7 label classes, and the softmax regression model is defined as follows:

$$y = \frac{1}{\sum_{k=1}^{n7} e^{W_{L,K}^T x_L + b_{L,K}}} \begin{bmatrix} e^{W_{L,1}^T x_L + b_{L,1}} \\ e^{W_{L,2}^T x_L + b_{L,2}} \\ \vdots \\ e^{W_{L,n7}^T x_L + b_{L,n7}} \end{bmatrix}$$

The output vector $y = x_{L+1}$ of the OUTPUT layer indicates the final probability of all classes of the current iteration.

**3.2.2 Backward propagation**
In the backward propagation phase, the drivable parameters are updated using the gradient descent method. It is achieved by minimizing a cost function and calculating the partial derivative of the cost function with respect to each parameter that can be driven [7]. The loss function used in this work is defined as:

$$J(\theta) = -m\frac{1}{m}\sum_{i=1}^{n1}\sum_{j=1}^{n7} 1\{j = Y^{(i)}\} \log(y_j^{(i)})$$

$$\delta_i = \frac{\partial J}{\partial u_i} = \begin{cases} -(Y-y) \circ f'(u_i), & i = L \\ (W_i^T \delta_{i+1} \circ f'(u_i), & i < L, \end{cases}$$

where ∘ denotes the multiplication element by element. $f'(u_i)$ can easily be represented as:

$$f'(u_i) = \begin{cases} (1-f(u_i)) \circ (1+f(u_i)), & i = 2,4,6 \\ 0 & i = 3,5 \\ f(u_i) \circ (1-f(u_i)), & i = 6 \end{cases}$$

Therefore, at each iteration, we would perform the update:
$$\theta = \theta - \alpha \nabla_\theta J(\theta)$$
to adjust the learning parameters, where is the learning factor, and

$$\nabla_\theta J(\theta) = \left\{ \frac{\partial J}{\partial \theta_1}, \frac{\partial J}{\partial \theta_2}, ..., \frac{\partial J}{\partial \theta_L} \right\}.$$

We know that $\theta$ contains $W_i$ and $b_i$, and

$$\frac{\partial J}{\partial \theta_i} = \left\{ \frac{\partial J}{\partial W_i}, \frac{\partial J}{\partial b_i} \right\},$$

where

$$\frac{\partial J}{\partial W_i} = \frac{\partial J}{\partial u_i} \circ \frac{\partial u_i}{\partial W_i} = \frac{\partial J}{\partial u_i} \circ x_i = \delta_i \circ x_i,$$

$$\frac{\partial J}{\partial b_i} = \frac{\partial J}{\partial u_i} \circ \frac{\partial u_i}{\partial b_i} = \frac{\partial J}{\partial u_i} = \delta_i$$

To obtain the best possible accuracy for these parameters, several tests were carried out. We implemented CNN with Keras and TensorFlow. We use a dropout of 0.1 in the two fully connected layers to avoid overlearning. The network parameters are learned over 96 iterations.

Fig. 3 show the red line indicating the training loss precision and the blue line is the validation loss precision.

The classification accuracy on the learning set is 97.39% and the validation accuracy is 97.77%. The learning error is 0.0754 and the validation error is 0.0655.

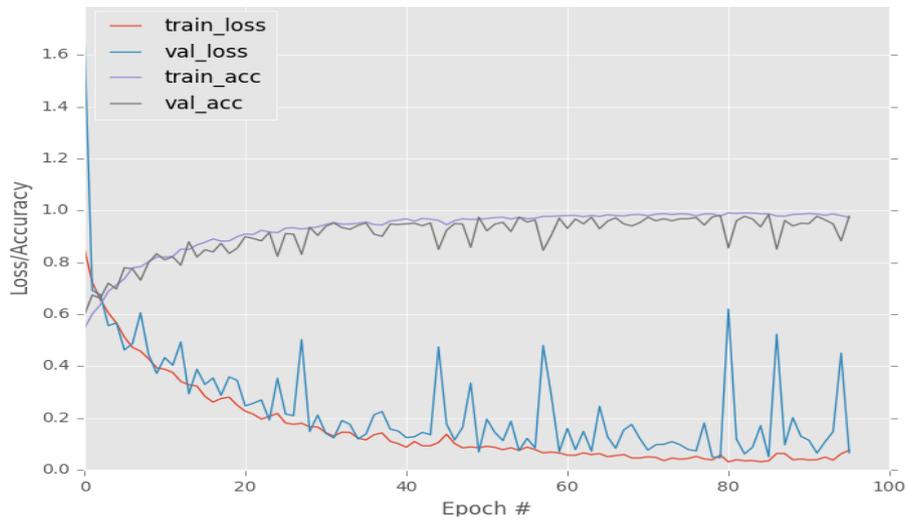

*Fig. 3: Performance of learning and validation*

## 4      Results

The classification accuracy on the test set is 95.3%. The test set consists of 213 images. Table 1 gives the confusion matrix for each class. There are 10 failures in recognition, which is mainly due to neutrophils and eosinophils. We can conclude that the parameters of our convolutional network model allow a good distinction in the classification between the different subtypes of blood.

Table 1: Confusion matrix

|  | EOSI. | LYMP. | MONO. | NEUTRO. |
|---|---|---|---|---|
| EOSI. | 42 | 0 | 0 | 3 |
| LYMP. | 0 | 63 | 0 | 0 |
| MONO. | 1 | 2 | 42 | 1 |
| NEUTRO. | 2 | 1 | 0 | 56 |

A test image is presented in the learning network parameter. The latter calculates the probability of detection of each blood subtype. The two highest probabilities are selected. The detected image is the one with the highest probability. In below fig.s (fig. 4 and fig. 5), we show the probabilities of detection for the four subtypes of blood.

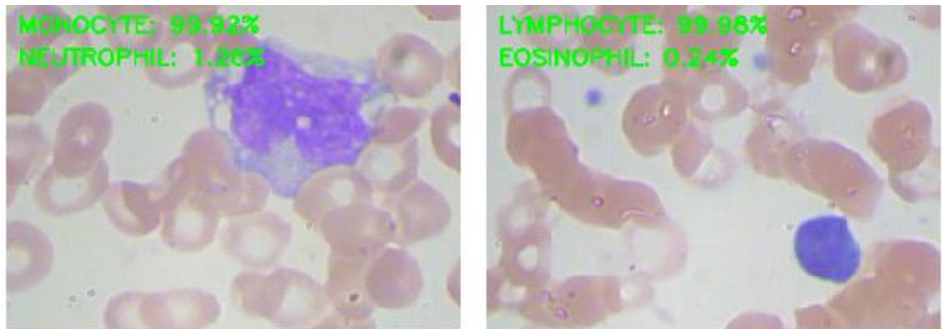

*Fig. 4: Test on a subtype of blood from the monocyte family (left) and the lymphocyte family (right)*

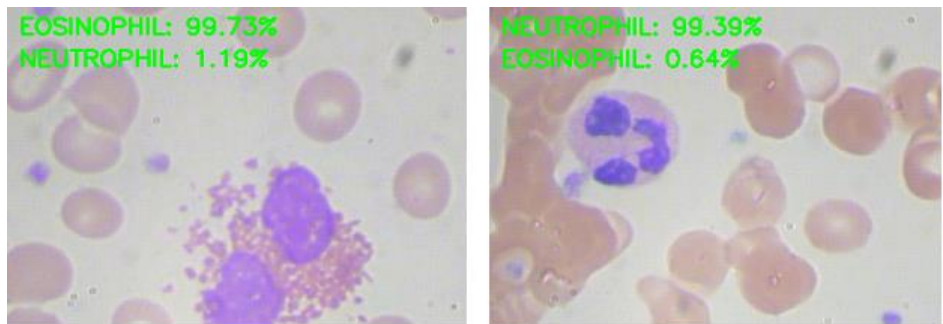

*Fig. 5: Test on a subtype of blood from the eosinophyll family (left) and the neutrophyl family (right).*

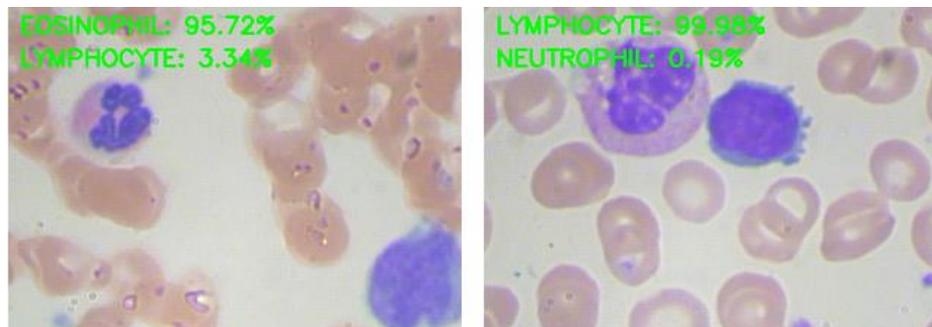

*Fig. 6: Neutrophyl and eosinophil subtypes falsely considered as eosinophil only (left) and monocyte and lymphocyte subtypes falsely considered as lymphocyte subtype only (right)*

The failures are to be found in the texture of the image. We can see that most failures is to be found in polynuclear where confusion between neutrophyl and eosinophil is noted. Another "failure" noted at the mononuclear level occurs mainly when the monocyte and lymphocyte subtypes are all present. The fig. 6 (left) shows the image as an eosinophil subtype whereas it is a neutrophil. Also a monocyte and lymphocyte subtypes labelled image is considered as 99.98% lymphocyte (right). That means, what we consider a failure is not totally one. Some images labeled as falsely detected show a sample of two blood subtypes. The CNN model detects the first right subtype only and ignores the second left subtype of the sample.

## 5    Conclusion

Based on convolutional neural networks, the deep learning algorithm we proposed was able to detect and distinguish the four subtypes of blood cells. Overall, the resulting confusion matrix is a good indicator of image detection and recognition capabilities. However, we note some detection failures especially for polynuclear subtypes such as neutrophil and eosinophil. With a test set of 213 images, only ten failures occurred.
We have noticed that the CNN model fails to detect more than two subtypes of blood in a sample at once, and that it would be interesting to propose a future multiple detection within a sample.

**ACKNOWLEDGMENT**
https://github.com/Shenggan/BCCD_Dataset MIT License

## References

1. D'Onofrio G, Zini G. Morphology of the blood. Stoneham, MA:Butterworth, Heinemann; 1998.
2. A. C. B. Monteiro, Y. Lano, R. P. França and N. Razmjooy, "chapter 2 WT-MO Algorithm: Automated Hematological Software Based on the Watershed Transform for Blood Cell Count", pp 41,  Book title "Applications of Image Processing and Soft Computing Systems in Agriculture", DOI: 10.4018/978-1-5225-8027-0.ch002, 2019
3. Sabino, D. M. U., da Fontoura Costa, L., Rizzatti, E. G., & Zago, M. A. (2004). A texture approach to leukocyte  recognition. Real-Time Imaging, 10(4),  205–216.
4. M. D. Joshi, A. H. Karode and S. R. Suralkar, (2013). "White Blood Cells Segmentation and Classification to Detect Acute Leukemia". International Journal of Emerging Trends & Technology in Computer Science (IJETTCS), 2(3): pp. 147-151.
5. S. Tantikitti,  S. Tumswadi and W. Premchaiswadi (2015). Image processing for detection of dengue virus based on WBC classification and decision tree. 13th International Conference on ICT and Knowledge Engineering, Bangkok, pp. 84-89.
6. M. Xu, D. P. Papageorgiou, S. Z. Abidi, M. Dao, H. Zhao, and G. E.  Karniadakis (2017).

"A deep convolutional neural network for classification of red blood cells in sickle cell anemia". PLoS Computational Biology, vol. 13 (10) : e1005746.
7. LeCun, Y., Boser, B., Denker, J., Henderson, D., Howard, R., Hubbard, W., and Jackel, L. (1990). "Handwritten Digit Recognition with a Back-Propagation Network". In Touretzky, D., editor, Advances in Neural Information Processing Systems, volume 2, pages 396-404, Denver 1989. Morgan Kaufmann, San Mateo
8. A.Krizhevsky, I. Sutskever, and G.E.Hinton, "Imagenet classification with deep convolutional neural networks," in Proceedings of the Advances in Neural Information Processing Systems 25 (NIPS '12), pp. 1097–1105, 2012.
9. Y. Lecun, L. Bottou, Y. Bengio and P. Haffner, Gradient-based learning applied to document recognition, in Proceedings of the IEEE, vol. 86, no. 11, pp. 2278-2324, Nov 1998.
10. D. H. Hubel and T. N. Wiesel, "Receptive fields and functional architecture of monkey striate cortex," The Journal of Physiology, vol. 195, no. 1, pp. 215–243, 1968.
11. Y. A. LeCun, L. Bottou, G. B. Orr, and K.-R. Müller, "Efficient backprop," in Neural Networks: Tricks of the Trade, pp. 9–48, Springer, Berlin, Germany, 2012.